\documentclass[aps,print,a4paper,showkeys,showpacs,twocolumn,10pt]{revtex4}%
\usepackage{amssymb}
\usepackage{amsmath}
\usepackage{graphicx}
\usepackage{amsfonts}%
\providecommand{\U}[1]{\protect \rule{.1in}{.1in}}
\begin{document}
\title{Nonautonomous solitons of Bose-Einstein condensation in a linear potential
with an arbitrary time-dependence}
\author{Qiu-Yan Li$^{1,2}$, Zai-Dong Li$^{1}$, Shu-Xin Wang$^{1}$, Wei-Wei Song$^{1}$,
Guangsheng Fu$^{2}$}
\affiliation{$^{1}$Department of Applied Physics, Hebei University of Technology, Tianjin
300401, China}
\affiliation{$^{2}$School of Information Engineer, Hebei University of Technology, Tianjin,
300401, China}

\begin{abstract}
In the presence of a linear potential with an arbitrary time-dependence,
Hirota method is developed carefully for applying into the effective
mean-field model of quasi-one-dimensional Bose-Einstein condensation with
repulsive interaction. We obtain the exact nonautonomous soliton solution
(NSS) analytically. These solutions show that the time-dependent potential can
affect the velocity of NSS. In some special cases the velocity has the
character of both increase and oscillation with time. A detail analysis for
the asymptotic behaviour of solutions shows that the collision of two NSSs is elastic.

\end{abstract}

\pacs{03.75.Lm, 05.30.Jp, 67.40.Fd}
\keywords{Nonautonomous soliton solution; interaction; Hirota method; Bose-Einstein condensation}\maketitle

\section{Introduction}

The concept of soliton was introduced firstly by Zabusky and Kruskal
\cite{Zabusky} to characterize nonlinear solitary waves that do not disperse
and completely preserve their localized form and speeds during propagation and
after a collision. This intrinsic favorable property of soliton has motived a
great of attention on the nonlinear systems in many fields of physics,
especially in high-rate telecommunications with optical fibers and condensate
physics. Hasegawa and Tappert \cite{Hasegawa} derived the nonlinear
Schr\"{o}dinger (NLS) equation model in fiber and firstly predicted optical
soliton, and then experimental verification has successfully been carried out
by Mollenauer et al. \cite{Mollenauer}. Since then, optical solitons have been
the objects of extensive theoretical and experimental studies in the past
three decades for their potential applications in long distance communication
and all-optical ultrafast switching devices. Recently, the controllable
soliton solutions \cite{Serkin,Serkin2,Serkin3,LiLu} are of interest in the
field of nonlinear optics and condensate physics, and then the term of
nonautonomous solitons \cite{Serkin3} introduced firstly. In fact, different
aspects of dynamics in nonautonomous models \cite{Chen,Konotop} in linear
potentials have been investigated theoretically. Strictly speaking, this
nonautonomous solutions obtained could not be considered as canonical
solitons. Fortunately, the realization of the Bose-Einstein condensation (BEC)
\cite{Anderson,Bradley} offered a good examples of the nonautonomous systems
in condensate physics.

With the realization of BECs the exploration of the nonlinear properties of
matter waves has been paid more particular interest. One of them is
macroscopically excited BECs, such as vortices \cite{Donnely} and solitons
\cite{Kivshar,Ruprecht,Zhang,Muryshev,Denschlag,Bronski,Carr}. At zero
temperature the dynamics of BEC is well described by the time-dependent
Gross-Pitaevskii (G-P) equation, and the nonlinearity results from the
interatomic interactions. Depending on the attractive or repulsive nature of
the interatomic interactions, G-P equation has of either bright or dark
soliton solutions, respectively. A bright soliton
\cite{Kevin,Khaykovich,Khawaja} in BEC is expected for the balance between the
dispersion and the attractive mean-field energy. However, large condensates
are necessarily associated with repulsive interaction, for which bright
soliton might seem impossible because the nonlinearity cannot compensate for
the kinetic energy part in the atomic dynamics. So it is interesting to
explore the property for dark soliton of BEC. A dark soliton \cite{Dum,Burger}
in BEC is a macroscopic excitation characterized by a local density minimum
and a phase gradient of the wave function at the position of the minimum.
Under the different conditions many soliton solutions
\cite{Zhang,Muryshev,Denschlag,Bronski,Carr,Shun,Lizd} have been obtained, as
well as the dynamics of the excitation of the condensate was discussed. When
the longitudinal dimension of BEC is much longer than its transverse dimension
which is the order of its healing length, the G-P equation can be reduced to
the quasi-one-dimensional (quasi-1D) regime. This trapped
quasi-one-dimensional \cite{Gorlitz} condensate has offered an useful tool to
investigate the nonlinear excitations such as solitons and vortices, which are
more stable than in 3D, where the solitons suffer from the transverse
instability and the vortices can bend. So the studies of both theory and
experiment are very important for the soliton excitation in
quasi-one-dimensional BEC.

The effective mean-field model of a quasi-1D BEC in a linear potential with an
arbitrary time-dependence is given by
\begin{equation}
i\hbar \frac{\partial}{\partial T}\Psi=-\frac{\hbar^{2}}{2m}\frac{\partial^{2}%
}{\partial X^{2}}\Psi+Xf(T)\Psi+g\left \vert \Psi \right \vert ^{2}\Psi,
\label{NLS}%
\end{equation}
where $%
{\displaystyle \int}
\left \vert \Psi \right \vert ^{2}dX=N$ is the number of atoms in the condensate.
The interacting constant of two-atom is $g$ $=2\hbar^{2}a/ml_{0}^{2}$
\cite{Petrov}, where $m$ is the mass of the atom, $a$ is the $s$-wave
scattering length ($a<0$ for attractive interaction; while $a>0$ for repulsive
interaction), and $l_{0}\equiv \sqrt{\hbar/m\omega_{0}}$ is the characteristic
extension length of the ground state wave function of harmonic oscillator. For
pithiness, we introduce $x=X/l_{0}$, $t=T/ml_{0}^{2}/\hbar$, and $\psi
=\Psi/\sqrt{Nl_{0}}$, and then Eq. (\ref{NLS}) reduces to the dimensionless
form
\begin{equation}
i\frac{\partial}{\partial t}\psi+\frac{1}{2}\frac{\partial^{2}}{\partial
x^{2}}\psi+xf\left(  t\right)  \psi-\mu \left \vert \psi \right \vert ^{2}\psi=0,
\label{NLS1}%
\end{equation}
where $\mu=2Nl_{0}a$, and $f\left(  t\right)  =-\frac{ml_{0}^{3}}{\hbar^{2}%
}f(\frac{T}{\hbar/ml_{0}^{2}})$. As in Ref. \cite{Serkin3}, Eq. (\ref{NLS1})
was called nonautonomous NLS model in linear potential. When $f\left(
t\right)  =$\textit{constant} and $a<0$, the Lax pair and NSSs in
inhomogeneous plasma has been constructed by the inverse scattering method
\cite{Chen}. F-expansion method \cite{ZhangJF} and Hirota method \cite{li2007}
were also developed to construct the bright NSSs in quasi-1D BECs. However,
the dynamic property of NSS of Eq. (\ref{NLS1}) hasn't well explored, and
Hirota method developed for Eq. (\ref{NLS1}) with the repulsive interaction is
also very interesting.

In the present paper we consider mainly the dynamics of dark nonautonomous
soliton which can be affected by adjusting the external linear time-dependent
potential. In the following section we demonstrate how to construct the exact
dark NSSs of Eq. (\ref{NLS1}), and the corresponding properties of such
solutions are studied in detail.

\section{Developed Hirota method and one Nonautonomous soliton solution}

Hirota method \cite{Hirota} is an effective straightforward technique to solve
the nonlinear equations. In order to clear the derivation of solution we
introduce the main idea of Hirota method briefly. Firstly, it apply a direct
transformation to the nonlinear equation. Then, by means of some skillful
bilinear operators the nonlinear equation can be decoupled into a series of
equations. With some reasonable assumptions the exact solutions can be
constructed effectively. However, in the presence of the time-dependent
potential the application of Hirota method should be more careful to get NSSs
of Eq. (\ref{NLS1}) in the case of the repulsive interaction.

Performing the normal procedure, we consider the complex function $G\left(
x,t\right)  $ and the real function $F\left(  x,t\right)  $ forming the
transformation%
\begin{equation}
\psi=\frac{G(x,t)}{F(x,t)}. \label{hirota1}%
\end{equation}
Substituting Eq. (\ref{hirota1}) into Eq. (\ref{NLS1}) we have%
\begin{equation}
F(iD_{t}+\frac{D_{x}^{2}}{2})G\cdot F+GF^{2}xf(t)-G(\frac{D_{x}^{2}}{2}F\cdot
F+\mu \overline{G}G)=0, \label{hirota2}%
\end{equation}
where the overbar denotes the complex conjugate, $D_{t}$ and $D_{x}^{2}$ are
called Hirota bilinear operators defined by
\begin{align}
&  D_{x}^{m}D_{t}^{n}G\left(  x,t\right)  \cdot F\left(  x^{\prime},t^{\prime
}\right) \label{b1}\\
&  =\left.  (\frac{\partial}{\partial x}-\frac{\partial}{\partial x^{\prime}%
})^{m}(\frac{\partial}{\partial t}-\frac{\partial}{\partial t^{\prime}}%
)^{n}G\left(  x,t\right)  F\left(  x^{\prime},t^{\prime}\right)  \right \vert
_{x=x^{\prime},t=t^{\prime}}.\nonumber
\end{align}
In the absence of the time-dependent potential, i.e., $xf(t)=0$, Eq.
(\ref{hirota2}) can be decoupled easily into two equations. In order to get
exact dark NSSs of Eq. (\ref{NLS1}), a real parameter $\lambda$ to be
determined should be added to Hirota bilinear operators in the presence of the
time-dependent potential, and then many attempts show that Eq. (\ref{hirota2})
can be decoupled into
\begin{equation}
\mathcal{\hat{A}}_{1}G\cdot F=0,\mathcal{\hat{A}}_{2}F\cdot F=-\mu
G\overline{G}, \label{hirota3}%
\end{equation}
where the overbar denotes the complex conjugate, and Hirota bilinear operators
$\mathcal{\hat{A}}_{1}$ and $\mathcal{\hat{A}}_{2}$ are given by
\begin{align*}
\mathcal{\hat{A}}_{1}  &  =iD_{t}+\frac{1}{2}D_{x}^{2}+xf(t)-\lambda,\\
\mathcal{\hat{A}}_{2}  &  =\frac{1}{2}D_{x}^{2}-\lambda.
\end{align*}
The derivation of Eq. (\ref{hirota3}) has made Eq. (\ref{NLS1}) into the
normal procedure of Hirota method. The spatial and time dependence term
$xf(t)$ will play an important role for getting the exact NSSs as shown later.

If the expressions of $G\left(  x,t\right)  $ and $F\left(  x,t\right)  $ are
obtained from Eq. (\ref{hirota3}), the exact dark NSSs can be expressed
analytically. For this purpose we assume that%
\begin{equation}
G=G_{0}\left(  1+\chi G_{1}\right)  ,F=\left(  1+\chi F_{1}\right)  ,
\label{ansaz1}%
\end{equation}
where $\chi$ is an arbitrary auxiliary parameter which will be absorbed in the
expression of NSSs. Substituting Eq. (\ref{ansaz1}) into Eq. (\ref{hirota3}),
and collecting the coefficients with same power of $\chi$, we have

(1) for the coefficient of $\chi^{0}$%
\begin{align}
\mathcal{\hat{A}}_{1}\left(  G_{0}\cdot1\right)   &  =0,\label{term1a}\\
\mathcal{\hat{A}}_{2}\left(  1\cdot1\right)   &  =-\mu G_{0}\overline{G}_{0},
\label{term1b}%
\end{align}

(2) for the coefficient of $\chi^{1}$%
\begin{align}
\mathcal{\hat{A}}_{1}\left(  G_{0}G_{1}\cdot1+G_{0}\cdot F_{1}\right)   &
=0,\label{term2a}\\
\mathcal{\hat{A}}_{2}\left(  F_{1}\cdot1+1\cdot F_{1}\right)   &  =-\mu
G_{0}\overline{G}_{0}\left(  G_{1}+\overline{G}_{1}\right)  , \label{term2b}%
\end{align}

(3) for the coefficient of $\chi^{2}$%
\begin{align}
\mathcal{\hat{A}}_{1}\left(  G_{0}G_{1}\cdot F_{1}\right)   &
=0,\label{term3a}\\
\mathcal{\hat{A}}_{2}\left(  F_{1}\cdot F_{1}\right)   &  =-\mu G_{0}%
\overline{G}_{0}G_{1}\overline{G}_{1}. \label{term3b}%
\end{align}

Using the definition of Hirota bilinear operator (\ref{b1}) the above
equations can be expressed in detail. Considering the presence of the term
$xf(t)$ in Eq. (\ref{term1a}) we assume $G_{0}$ has the form%
\begin{equation}
G_{0}=\gamma_{0}e^{i\eta_{0}}, \label{solution1}%
\end{equation}
where
\begin{equation}
\eta_{0}=P_{0}(t)x+\Omega_{0}(t), \label{para0}%
\end{equation}
with $P_{0}(t)$ and $\Omega_{0}(t)$ is to be determined, respectively.
Substituting $G_{0}$ into Eq. (\ref{term1a}) we have%
\[
\text{\ }0=\left[  -P_{0,t}(t)+f(t)\right]  x-\Omega_{0,t}(t)-\frac{1}{2}%
P_{0}^{2}(t)-\lambda,
\]
which implies the solution%
\begin{align}
P_{0}\left(  t\right)   &  =\int \nolimits_{0}^{t}f\left(  \tau \right)
d\tau+\xi_{0},\text{ }\nonumber \\
\Omega_{0}\left(  t\right)   &  =-\frac{1}{2}\int \nolimits_{0}^{t}P_{0}%
^{2}\left(  \tau \right)  d\tau-\lambda t+\zeta_{0}, \label{para0a}%
\end{align}
where $\xi_{0}$ and $\zeta_{0}$ is an arbitrary real constant, respectively.
From the restriction of Eq. (\ref{term1b}) we get $\left \vert \gamma
_{0}\right \vert ^{2}=\lambda/\mu$ which shows that the exist of dark NSS
demand the parameter $\lambda>0$. For convenience $\gamma_{0}$ can be chosen
as $\gamma_{0}=\sqrt{\lambda/\mu}$.

Expanding Eqs. (\ref{term2a}) and (\ref{term2b}) with the definition of Eq.
(\ref{b1}) one can see that $G_{1}$ and $F_{1}$ admit the expression
\begin{equation}
G_{1}=Z_{1}\exp \eta_{1},F_{1}=\exp \eta_{1}, \label{solution1a}%
\end{equation}
where the parameter $Z_{1}$ to be determined is complex, and the real
parameter $\eta_{1}$ is given by%
\begin{equation}
\eta_{1}=P_{1}\left(  t\right)  x+\Omega_{1}\left(  t\right)  . \label{para1}%
\end{equation}
Substituting Eqs. (\ref{solution1}) and (\ref{solution1a}) into Eq.
(\ref{term2a}) we have%
\begin{equation}
0=i\left(  Z_{1}-1\right)  \left(  P_{1,t}x+\Omega_{1,t}+P_{0}P_{1}\right)
+\frac{1}{2}\left(  Z_{1}+1\right)  P_{1}^{2}. \label{pa1}%
\end{equation}
In the case of $Z_{1}\neq \pm1$(the case of $Z_{1}=\pm1$ will be discussed in
below), the solution of Eq. (\ref{pa1}) is $P_{1,t}=0$, i.e., $P_{1}$ should
be independent on $t$, and $Z_{1}$ is given by%
\begin{equation}
Z_{1}=\frac{iP_{0}P_{1}-\frac{1}{2}P_{1}^{2}+i\Omega_{1,t}}{iP_{0}P_{1}%
+\frac{1}{2}P_{1}^{2}+i\Omega_{1,t}}, \label{z1}%
\end{equation}
which shows that $\left \vert Z_{1}\right \vert ^{2}=1$. From Eqs.
(\ref{term2b}) to (\ref{term3b}) we get the restriction%
\begin{equation}
P_{1}^{2}=2\lambda-\lambda \left(  Z_{1}+\overline{Z}_{1}\right)  . \label{p1}%
\end{equation}
Then from Eq. (\ref{z1}) and Eq. (\ref{p1}) we obtain
\begin{align}
Z_{1}  &  =\frac{\sqrt{4\lambda-P_{1}^{2}}+iP_{1}}{\sqrt{4\lambda-P_{1}^{2}%
}-iP_{1}},\nonumber \\
\Omega_{1}  &  =-P_{1}\int \nolimits_{0}^{t}P_{0}\left(  \tau \right)
d\tau+\frac{1}{2}\sqrt{4\lambda-P_{1}^{2}}P_{1}t+\zeta_{1}, \label{para1a}%
\end{align}
where $\zeta_{1}$ is a real constant, $P_{1}$ is an arbitrary real parameter,
and $\lambda \geq P_{1}^{2}/4$.

With the help of Eqs. (\ref{ansaz1}), (\ref{solution1}) and (\ref{solution1a}%
), after absorbing $\chi$, the exact dark NSS of Eq. (\ref{NLS1}) can be
derived as\
\begin{equation}
\psi_{1}=\frac{1}{2}\sqrt{\frac{\lambda}{\mu}}e^{i\eta_{0}}\left[  \left(
1+Z_{1}\right)  -\left(  1-Z_{1}\right)  \tanh \frac{\eta_{1}}{2}\right]  ,
\label{onesoliton}%
\end{equation}
where the parameters $\eta_{0}$, $\eta_{1}$, and $Z_{1}$ have been given in
Eqs. (\ref{para0}), (\ref{para0a}), (\ref{para1}) and (\ref{para1a}).

If one chose $\lambda=4\eta^{2}$ and $P_{1}=4a_{1}\eta^{2}$, the solution in
Eq. (\ref{onesoliton}) has the same form as the general solution obtained in
the framework \cite{Serkin3}. In the case of $f(t)=0$, the solution in Eq.
(\ref{onesoliton}) reduces to the one soliton solution of the normal NLS
equation. When $f(t)=$\textit{constant}, the solution (\ref{onesoliton})
represents the nonlinear dark wave propagation in linearly inhomogeneous
plasma \cite{Chen} or optical fibre with the abnormal dispersion. As
$f(t)=b_{1}+l\cos \left(  \omega t\right)  $, the solution (\ref{onesoliton})
denotes NSS in BEC with considering the coupling of the external field and the
effect of gravity.

From the NSS in Eq. (\ref{onesoliton}) we clear two special case mentioned
before, i.e., $Z_{1}=\pm1$. When $Z_{1}=1$, the solution (\ref{onesoliton})
reduces to plane-wave solution $\psi_{1}=\sqrt{\lambda/\mu}e^{i\eta_{0}}$,
which corresponds to the uniform distribution density of bosons. On the other
hand, when $Z_{1}=-1$ the solution (\ref{onesoliton}) becomes $\psi_{1}%
=-\sqrt{\lambda/\mu}e^{i\eta_{0}}\tanh \left(  \eta_{1}/2\right)  $, where
$\eta_{1}=2\sqrt{\lambda}x-2\sqrt{\lambda}\int \nolimits_{0}^{t}P_{0}\left(
\tau \right)  d\tau+\zeta_{1}$. This solution represents the black NSS of BEC
in a linear potential with an arbitrary time-dependence which is caused by the
coupling of external field and the effect of gravity.

We also find the effect of term $xf(t)$ from the NSS in Eq. (\ref{onesoliton}%
). As shown in the expression of $\eta_{0}$, the term $xf(t)$ can only
contribute a phase to the background. The width of nonautonomous soliton,
defined by $1/P_{1}$, is not affected by the time-dependent external
potential. From Eqs. (\ref{para1}) and (\ref{para1a}) we get the velocity
\[
V_{1}=-\frac{\partial}{\partial t}\frac{\Omega_{1}}{P_{1}}=\int \nolimits_{0}%
^{t}f\left(  \tau \right)  d\tau+\xi_{0}-\frac{1}{2}\sqrt{4\lambda-P_{1}^{2}},
\]
which shows that the time-dependent potential play an important role for the
velocity of dark NSS. When $f(t)=b_{1}+l\cos \left(  \omega t\right)  $, the
velocity becomes $V_{1}=b_{1}t+l/\omega \sin \left(  \omega t\right)  +\xi
_{0}-\sqrt{\lambda-\left(  P_{1}/2\right)  ^{2}}$, which increases and
oscillates with time. These results shows that under the effect of gravity,
the BEC slides down where the dark nonautonomous soliton slides down and
oscillates in time. It can be realized by controlling the external field.

\section{Collision of Two dark Nonautonomous Solitons}

In this section we will give the analytical expression of two dark NSSs of Eq.
(\ref{NLS1}). In this case $G\left(  x,t\right)  $ and $F\left(  x,t\right)  $
of Eq. (\ref{hirota1}) are assumed as%
\begin{equation}
G=G_{0}\left(  1+\chi G_{1}+\chi^{2}G_{2}\right)  ,F=1+\chi F_{1}+\chi
^{2}F_{2}. \label{ansaz2}%
\end{equation}
where $G_{0}$ has been obtained in Eq. (\ref{solution1}). Employing the
similar procedure of the above section we obtain the following set of
equations from Eq. (\ref{hirota3})

(1) for the coefficient of $\chi$%
\begin{align}
\pounds _{1}\left(  G_{1}\cdot1+1\cdot F_{1}\right)   &  =0,\label{th1a}\\
\mathcal{\hat{A}}_{2}\left(  1\cdot F_{1}+F_{1}\cdot1\right)   &
=-\lambda \left(  G_{1}+\overline{G}_{1}\right)  , \label{th1b}%
\end{align}

(2) for the coefficient of $\chi^{2}$%
\begin{align}
\pounds _{1}\left(  1\cdot F_{2}+G_{1}\cdot F_{1}+G_{2}\cdot1\right)   &
=0,\label{th2a}\\
\mathcal{\hat{A}}_{2}\left(  1\cdot F_{2}+F_{1}\cdot F_{1}+F_{2}\cdot1\right)
&  =-\lambda \left(  G_{2}+\overline{G}_{1}G_{1}+\overline{G}_{2}\right)  ,
\label{th2b}%
\end{align}

(3) for the coefficient of $\chi^{3}$%
\begin{align}
\pounds _{1}\left(  G_{1}\cdot F_{2}+G_{2}\cdot F_{1}\right)   &
=0,\label{th3a}\\
\mathcal{\hat{A}}_{2}\left(  F_{1}\cdot F_{2}+F_{2}\cdot F_{1}\right)   &
=-\lambda \left(  \overline{G}_{2}G_{1}+\overline{G}_{1}G_{2}\right)  ,
\label{th3b}%
\end{align}

(4) for the coefficient of $\chi^{4}$%
\begin{align}
\pounds _{1}G_{2}\cdot F_{2}  &  =0,\label{th4a}\\
\mathcal{\hat{A}}_{2}F_{2}\cdot F_{2}  &  =-\lambda \overline{G}_{2}G_{2},
\label{th4b}%
\end{align}
where $\pounds _{1}=iD_{t}+\frac{1}{2}D_{x}^{2}+P_{0}(t)D_{x},$ and
$\mathcal{\hat{A}}_{2}$ is given before.

It is obvious that one can solve the equations (\ref{th1a}) to (\ref{th4b}) in
turn with the reasonable expressions of $G_{1}$ and $F_{1}$. A detail analysis
shows that $G_{1}$ and $F_{1}$ admit the forms%
\begin{equation}
G_{1}=Z_{1}\exp \eta_{1}+Z_{2}\exp \eta_{2},F_{1}=\exp \eta_{1}+\exp \eta_{2},
\label{Two1}%
\end{equation}
where $Z_{j}$ is complex and $\eta_{j}$ has the form
\begin{equation}
\eta_{j}=P_{j}\left(  t\right)  x+\Omega_{j}\left(  t\right)  ,j=1,2,
\label{para2}%
\end{equation}
with the parameter $P_{j}(t)$ and $\Omega_{j}(t)$ is to be determined,
respectively. Substituting Eq. (\ref{Two1}) into Eq. (\ref{th1a}) we have%
\begin{align*}
0  &  =e^{\eta_{1}}[(Z_{1}-1)(iP_{1,t}x+iP_{0}P_{1}+i\Omega_{1,t})+\frac
{Z_{1}+1}{2}P_{1}^{2}]\\
&  +e^{\eta_{2}}[(Z_{2}-1)(iP_{2,t}x+iP_{0}P_{2}+i\Omega_{2,t})+\frac{Z_{2}%
+1}{2}P_{2}^{2}].
\end{align*}
In the case of $Z_{1}\neq \pm1$ and $Z_{2}\neq \pm1$, the above equation implies
that $P_{j,t}\left(  t\right)  =0,$ $j=1,2,$ i.e., $P_{j}$ is independent on
$t$, and $Z_{j}$ is given by%
\begin{equation}
Z_{j}=\frac{iP_{0}P_{j}-\frac{1}{2}P_{j}^{2}+i\Omega_{j,t}}{iP_{0}P_{j}%
+\frac{1}{2}P_{j}^{2}+i\Omega_{j,t}}, \label{zj}%
\end{equation}
which shows that $\left \vert Z_{j}\right \vert ^{2}=1$, $j=1,2$. From Eq.
(\ref{th1b}) we have the restriction%
\begin{equation}
P_{j}^{2}=2\lambda-\lambda \left(  Z_{j}+\overline{Z}_{j}\right)  ,j=1,2.
\label{pj}%
\end{equation}
From Eqs. (\ref{zj}) and (\ref{pj}) we have%
\begin{align}
Z_{j}  &  =\frac{\sqrt{4\lambda-P_{j}^{2}}+iP_{j}}{\sqrt{4\lambda-P_{j}^{2}%
}-iP_{j}},\nonumber \\
\Omega_{j}  &  =-P_{j}\int \nolimits_{0}^{t}P_{0}\left(  \tau \right)
d\tau+\frac{1}{2}\sqrt{4\lambda-P_{j}^{2}}P_{j}t+\zeta_{j}, \label{para2a}%
\end{align}
where $\zeta_{j}$ is a real constant, $P_{j}$ is an arbitrary real parameter,
and $\lambda \geq P_{j}^{2}/4$, $j=1,2$.

Substituting Eq. (\ref{Two1}) into Eqs. (\ref{th2a}) and (\ref{th2b}), after a
tedious and expatiatory calculation we obtain the expressions of $G_{2}$ and
$F_{2}$ as%
\begin{equation}
G_{2}=A_{12}Z_{1}Z_{2}\exp \left(  \eta_{1}+\eta_{2}\right)  ,F_{2}=A_{12}%
\exp \left(  \eta_{1}+\eta_{2}\right)  , \label{Two2}%
\end{equation}
where the real parameter $A_{12}$ is given by$\allowbreak$ $\allowbreak$
$\allowbreak$ $\allowbreak \allowbreak$%
\[
A_{12}=\frac{4\lambda-P_{1}P_{2}-\sqrt{4\lambda-P_{1}^{2}}\sqrt{4\lambda
-P_{2}^{2}}}{4\lambda+P_{1}P_{2}-\sqrt{4\lambda-P_{1}^{2}}\sqrt{4\lambda
-P_{2}^{2}}}.
\]
$\allowbreak \allowbreak$Now we have obtained the expression of $G_{0}%
,G_{1},G_{2},F_{1}$ and $F_{2}$ in Eq. (\ref{ansaz2}). With the help of Eqs.
(\ref{Two1}) to (\ref{Two2}) one can find the Eqs. (\ref{th3a}) to
(\ref{th4b}) are satisfied to the moment.

With Eqs. (\ref{hirota1}), (\ref{solution1}), (\ref{Two1}), and (\ref{Two2}),
while absorbing the parameter $\chi$, we obtain the dark two NSS of Eq.
(\ref{NLS1}) as%
\begin{equation}
\psi_{2}=\sqrt{\frac{\lambda}{\mu}}e^{i\eta_{0}}\frac{1+Z_{1}e^{\eta_{1}%
}+Z_{2}e^{\eta_{2}}+A_{12}Z_{1}Z_{2}e^{\eta_{1}+\eta_{2}}}{1+e^{\eta_{1}%
}+e^{\eta_{2}}+A_{12}e^{\eta_{1}+\eta_{2}}}. \label{twosoliton}%
\end{equation}
When $f(t)=0$, the solution (\ref{twosoliton}) denotes dark two solitons
interaction of the normal NLS equation. When $f(t)=$\textit{constant}, the
solution (\ref{twosoliton}) represents the dynamics of two nonautonomous
nonlinear waves in linearly inhomogeneous plasma or optical fibre with the
abnormal dispersion. As shown before, the expressions (\ref{onesoliton}) and
(\ref{twosoliton}) imply that Hirota method has more advantage for getting
such solutions as well.

The solution in Eq. (\ref{twosoliton}) describes a general scattering process
of two dark NSSs with different center velocity $V_{1}$ and $V_{2}$,
respectively. From Eqs. (\ref{para2}) and (\ref{para2a}) we get each velocity
as
\[
V_{j}=\int \nolimits_{0}^{t}f\left(  \tau \right)  d\tau+\xi_{0}-\frac{1}%
{2}\sqrt{4\lambda-P_{j}^{2}},j=1,2.
\]
Under the proper parameters two NSSs can move toward each other, one with the
velocity $V_{1}$, while the other with $V_{2}$. In order to understand the
nature of two NSSs interaction, we analyze the asymptotic behave of the
solution in Eq. (\ref{twosoliton}). Asymptotically, the solution in Eq.
(\ref{twosoliton}) can be written as a combination of two NSSs in Eq.
(\ref{onesoliton}). The asymptotic form of two NSSs in limits $t\rightarrow
-\infty$ and $t\rightarrow \infty$ is similar to that of the one NSS in Eq.
(\ref{onesoliton}).

(i) Before collision (limit $t\rightarrow-\infty$)

(a) Nonautonomous soliton 1 ($\eta_{1}\approx0$, $\eta_{2}\rightarrow-\infty
$)
\begin{equation}
\psi_{2}\rightarrow \frac{1}{2}\sqrt{\frac{\lambda}{\mu}}e^{i\eta_{0}}\left[
\left(  1+Z_{1}\right)  -\left(  1-Z_{1}\right)  \tanh \frac{\eta_{1}}%
{2}\right]  , \label{asym1}%
\end{equation}

(b) Nonautonomous soliton 2 ( $\eta_{2}\approx0$, $\eta_{1}\rightarrow \infty$)%
\begin{equation}
\psi_{2}\rightarrow \frac{1}{2}\sqrt{\frac{\lambda}{\mu}}Z_{1}e^{i\eta_{0}%
}\left[  \left(  1+Z_{2}\right)  -\left(  1-Z_{2}\right)  \tanh \frac{1}%
{2}\left(  \eta_{2}+\delta_{0}\right)  \right]  . \label{asym2}%
\end{equation}

(ii) After collision (limit $t\rightarrow \infty$)

(a) Nonautonomous soliton 1 ($\eta_{1}\approx0$, $\eta_{2}\rightarrow \infty$)%
\begin{equation}
\psi_{2}\rightarrow \frac{1}{2}\sqrt{\frac{\lambda}{\mu}}Z_{2}e^{i\eta_{0}%
}\left[  \left(  1+Z_{1}\right)  -\left(  1-Z_{1}\right)  \tanh \frac{1}%
{2}\left(  \eta_{1}+\delta_{0}\right)  \right]  , \label{asym3}%
\end{equation}

(b) Nonautonomous soliton 2 ( $\eta_{2}\approx0$, $\eta_{1}\rightarrow-\infty
$)%
\begin{equation}
\psi_{2}\rightarrow \frac{1}{2}\sqrt{\frac{\lambda}{\mu}}e^{i\eta_{0}}\left[
\left(  1+Z_{2}\right)  -\left(  1-Z_{2}\right)  \tanh \frac{\eta_{1}}%
{2}\right]  , \label{asym4}%
\end{equation}
where the center shift is given by $\delta_{0}=\ln A_{12}$. By analyzing the
asymptotic behave in detail we know that there is no change of the amplitude
for each NSS during collision, while one should notice that the factor
$\left \vert Z_{j}\right \vert =1$, $j=1,2$, again. However, from Eq.
(\ref{asym1}) to Eq. (\ref{asym4}) we find a phase exchange $\delta_{0}$ for
soliton 1 and soliton 2 during collision. These results show that the
collision of two NSSs is elastic.

\section{Conclusion}

In this paper, we report the exact dark NSSs of quasi-one-dimensional BEC in a
linear potential with an arbitrary time-dependence, and Hirota method is also
developed. With the skillful assumption the exact dark NSSs are constructed
effectively. From these results we find the time-dependent potential can
affect the velocity of NSS. In some special cases the velocity of NSS in
quasi-one-dimensional BEC increases and oscillations in time, the BEC slides
down under the effect of gravity. We also investigate the asymptotic behave of
two NSSs which denotes the elastic collision.

\section{Acknowledgement}

This work is supported by NSF of China under Grant No. 10874038, the Natural
Science Foundation of Hebei Province of China under Grant No. A2008000006, and
the key subject construction project of Hebei Provincial University of China.


\begin{thebibliography}{99}                                                                                               %


\bibitem {Zabusky}N. J. Zabusky, and M. D. Kruskal, Phys. Rev. Lett. 15 (1965) 240.

\bibitem {Hasegawa}A. Hasegawa, F. Tappert, Appl. Phys. Lett. 23 (1973) 142.

\bibitem {Mollenauer}L.F. Mollenauer, R.H. Stolen, J.P. Gordon, Phys. Rev.
Lett. 45 (1980) 1095.

\bibitem {Serkin}V. N. Serkin and A. Hasegawa, Phys. Rev. Lett. 85 (2000) 4502.

\bibitem {Serkin2}V. N. Serkin, et al., JETP Lett. 74 (2000) 573;

V. N. Serkin, JETP Lett. 72 (2000) 89;

V. N. Serkin, et al., IEEE J. Quantum Electron. 8 (2002) 418.

\bibitem {Serkin3}V. N. Serkin, A. Hasegawa, and T. L. Belyaeva, Phys. Rev.
Lett. 98 (2007) 074102.

\bibitem {LiLu}Lei Wu, Jie-Fang Zhang, Lu Li, C. Finot, and K. Porsezian,
Phys. Rev. A 78 (2008) 053807

\bibitem {Chen}H. H. Chen and C. S. Liu, Phys. Rev. Lett. 37 (1976) 693; Phys.
Fluids 21 (1978) 377

\bibitem {Konotop}V. V. Konotop, Phys. Rev. E 47 (1993) 1423;

V. V. Konotop, O. A. Chubykalo, and L. V\'{a}zquez, Phys. Rev. E 48 (1993) 563.

V. V. Konotop, Theoretical and Mathematical Physics, 99 (1994) 687.

\bibitem {Anderson}M. H. Anderson, J. R. Ensher, M. R. Matthews, C. E. Wieman,
and E. A. Cornell, Science 269 (1995) 198;

K. B. Davis, M.-O. Mewes, M. R. Andrews, N. J. van Druten, D. S. Durfee, D. M.
Kurn, and W. Ketterle, Phys. Rev. Lett. 75 (1995) 3969.

\bibitem {Bradley}C. C. Bradley, C. A. Sackett, J. J. Tollett, and R. G.
Hulet, Phys. Rev. Lett. 75 (1995) 1687; 78 (1997) 985;

D. G. Fried, T. C. Killian, L. Willmann, D. Landhuis, S. C. Moss, D. Kleppner,
and T. J. Greytak, Phys. Rev. Lett. 81 (1998) 3811.

\bibitem {Donnely}R. J. Donnely, Quantized Vortices in Helium II (Cambridge
University Press, Cambridge, 1991).

M. R. Matthews, B. P. Anderson, P. C. Haljan, D. S. Hall, C. E. Wieman, and E.
A. Cornell, Phys. Rev. Lett. 83 (1999) 2498.

A. L. Fetter and A. A. Svidzinsky, J. Phys.: Condens. Matter 13 (2001) R135.

\bibitem {Kivshar}Y. S. Kivshar and B. Luther-Davies, Phys. Rep. 298 (1998) 81.

\bibitem {Ruprecht}P. A. Ruprecht, M. J. Holland, K. Burnett, and M. Edwards,
Phys. Rev. A 51 (1995) 4704.

\bibitem {Zhang}W. Zhang, D. F. Walls, and B. C. Sanders, Phys. Rev. Lett. 72
(1994) 60;

W. P. Reinhardt and C. W. Clark, J. Phys. B 30 (1997) L785;

A. D. Jackson, G. M. Kavoulakis, and C. J. Pethick, Phys. Rev. A 58 (1998) 2417.

\bibitem {Muryshev}A. E. Muryshev, H. B. van Linden van den Heuvell, and G.V.
Shlyapnikov, Phys. Rev. A 60 (1999) R2665;

Lincoln D Carr, Mary Ann Leung, and William P Reinhardt, J. Phys. B 33 (2000) 3983.

\bibitem {Denschlag}J. Denschlag, et al., Science 287 (2000) 97.

\bibitem {Bronski}J. C. Bronski, L. D. Carr, B. Deconinck, and J. N. Kutz,
Phys. Rev. Lett. 86 (2001) 1402.

\bibitem {Carr}L. D. Carr, C. W. Clark, and W. P. Reinhardt, Phys. Rev. A 62
(2000) 063610; 62 (2000) 063611;

D. L. Feder et al., Phys. Rev. A 62 (2000) 053606.

\bibitem {Kevin}Kevin E. Strecker, Guthrie B. Partridge, Andrew G. Truscott,
and Randall G. Hulet, Nature (London) 417 (2002) 150.

\bibitem {Khaykovich}L. Khaykovich, F. Schreck, G. Ferrari, T. Bourdel, J.
Cubizolles, L. D. Carr, Y. Castin, and C. Salomon, Science 296 (2002) 1290.

\bibitem {Khawaja}U. Al Khawaja, H. T. C. Stoof, R. G. Hulet, K. E. Strecker,
and G. B. Partridge, Phys. Rev. Lett. 89 (2002) 200404.

\bibitem {Dum}R. Dum, J. I. Cirac, M. Lewenstein, and P. Zoller, Phys. Rev.
Lett. 80 (1998) 2972.

\bibitem {Burger}S. Burger, K. Bongs, S. Dettmer, W. Ertmer, and K. Sengstock,
Phys. Rev. Lett. 83 (1999) 5198.

\bibitem {Shun}Shun-Jin Wang, Cheng-Long Jia, Dun Zhao, Hong-Gang Luo, and
Jun-Hong An, Phys. Rev. A 68 (2003) 015601.

\bibitem {Lizd}Zai-Dong Li, P. B. He, Lu Li, J.-Q. Liang, and W. M. Liu, Phys.
Rev. A 71 (2005) 053611;

Lu Li, Zaidong Li, Boris A. Malomed, Dumitru Mihalache, and W. M. Liu, Phys.
Rev. A 72 (2005) 033611;

Q. Y. Li, Z. W. Xie, L. Li, Z. D. Li, and J. Q. Liang, Annals of Physics 312
(2004) 128.

\bibitem {Gorlitz}A. G\"{o}rlitz, J. M. Vogels, A. E. Leanhardt, C. Raman, T.
L. Gustavson, and J. R. Abo-Shaeer, Phys. Rev. Lett. 87 (2001) 130402;

S. Dettmer et al., ibid. 87 (2001) 160406.

\bibitem {Petrov}D. S. Petrov, G.V. Shlyapnikov, and J. T. M. Walraven, Phys.
Rev. Lett. 85 (2000) 3745.

\bibitem {ZhangJF}Qin Yang, and Jie-fang Zhang, Optics Communications 258
(2006) 35.

\bibitem {li2007}Zai-Dong Li, Qiu-Yan Li, Xing-Hua Hu, Zhong-Xi Zheng, Yubao
Sun, Ann. Phys. (N.Y.) 322 (2007) 2545.

\bibitem {Hirota}R. Hirota, J. Phys. Soc. Jpn. 51 (1982) 323.
\end{thebibliography}
\end{document}